\documentclass[useAMS,usenatbib]{mn2e}

\usepackage{longtable}
\usepackage{pdflscape}
\usepackage{graphicx}
\usepackage{pdflscape}
\usepackage{lscape,graphicx,pifont}
\usepackage{color}

\newcommand{\Msun}{\mbox{\,M$_\odot$}}
\newcommand{\Rsun}{\mbox{\,R$_\odot$}}

\newcommand{\vunit}{\mbox{\,km\,s$^{-1}$}}
\newcommand{\mic}{\mbox{$\,\mu$m}}
\newcommand{\pion}[2]{{#1}\,{\sc {#2}}}
\newcommand{\fion}[2]{[{#1}\,{\sc {#2}}]}
\newcommand{\nucl}[2]{\mbox{$^{#1}${#2}}}

\newcommand{\ltsimeq}{\raisebox{-0.6ex}{$\,\stackrel
        {\raisebox{-.2ex}{$\textstyle <$}}{\sim}\,$}}
\newcommand{\gtsimeq}{\raisebox{-0.6ex}{$\,\stackrel
        {\raisebox{-.2ex}{$\textstyle >$}}{\sim}\,$}}

\newcommand{\rnsgr}{\mbox{V3890~Sgr}}

\newcommand{\sirtf}{\mbox{\it Spitzer}}
\newcommand{\spitzer}{\mbox{\it Spitzer Space Telescope}}

\title[The red giant in V3890~Sgr]{The recurrent nova V3890~Sgr:
a near-infrared and optical study of 
the red giant component and its environment}

\author[B. Kaminsky et al.]{B. Kaminsky$^{1}$\thanks{E-mail: bogdan@mao.kiev.ua},
A. Evans$^{2}$\thanks{Corresponding author. Email: a.evans@keele.ac.uk},
Ya. V. Pavlenko$^{1,3,4}$,
C. E. Woodward$^5$\thanks{Visiting Astronomer at the Infrared
Telescope Facility, which is operated by the University of Hawaii under contract
80HQTR19D0030 with the National Aeronautics and Space Administration.},  \newauthor
D. P. K. Banerjee$^6$,
R. D. Gehrz$^5$, 
F. Walter$^7$,
S. Starrfield$^8$,
I. Ilyin$^9$, \newauthor
K. G. Strassmeier$^9$,
R. M. Wagner$^{10,11}$
 \\
$^{1}$Main Astronomical Observatory, Academy of Sciences of the Ukraine, 
Golosiiv Woods, Kyiv-127, 03680 Ukraine\\ 
${^2}$Astrophysics Group, Lennard Jones Laboratory, Keele University, Keele,
Staffordshire. ST5 5BG, UK\\
${^3}$Centre for Astrophysics Research, University of Hertfordshire, College Lane, 
Hatfield, AL10 9AB, UK \\
$^4$Nicolaus Copernicus Astronomical Center, ul. Rabianska 8, 87-100 Toru\'n, Poland \\
$^5$Minnesota Institute for Astrophysics, School of Physics \& Astronomy,
116 Church Street SE, University of Minnesota, \\
Minneapolis, MN 55455, USA\\ 
$^{6}$Physical Research Laboratory, Navrangpura,  Ahmedabad, Gujarat 
380009, India \\ 
$^7$Department of Physics \& Astronomy, Stony Brook University, 
Stony Brook, NY, 11794-3800, USA \\ 
$^8$School of Earth and Space Exploration, Arizona State University, 
Box 871404, Tempe, AZ 85287-1404, USA\\ 
$^9$Leibniz-Institut f\"ur Astrophysik Potsdam (AIP), An der Sternwarte 16,
D-14482 Potsdam, Germany\\ 
$^{10}$Department of Astronomy, The Ohio State University, 140 W. 18th Avenue, Columbus, OH 43210, USA \\ 
$^{11}$Large Binocular Telescope Observatory, 933 North Cherry Avenue, Tucson, AZ 85721, USA }

\begin{document}

\date{Version of \today}

\pagerange{\pageref{firstpage}--\pageref{lastpage}} \pubyear{2021}

\maketitle

\label{firstpage}

\begin{abstract}
We present an analysis of the red giant component of the recurrent
nova \rnsgr, using data obtained before and after its 2019 eruption.
Its effective temperature is $T_{\rm eff}=3050\pm$200~K for 
$\log{g}=0.7$, although there are modest changes in $T_{\rm eff}$. 
There is an overabundance of both carbon ($0.20\pm0.05$~dex) and sodium 
($1.0\pm0.3$~dex) relative to their solar values, possibly the 
result of ejecta from the 1990 nova eruption being entrained into 
the red giant photosphere. We find \nucl{12}{C}/\nucl{13}{C} $=25\pm2$, 
a value similar to that found in red giants in other recurrent novae. 
The interpretation of the quiescent spectrum in the 
5--38\mic\ region requires the presence of photospheric SiO 
absorption and cool ($\sim400$~K) dust in the red giant environment.
The spectrum in the region of the \pion{Na}{i} D lines is complex,
and includes at least six interstellar components, together with 
likely evidence for interaction between ejecta from the 2019 eruption
and  material accumulated in the plane of the binary.
Three recurrent novae with giant secondaries have been shown to have
environments with different dust content, 
but photospheres with similar \nucl{12}{C}/\nucl{13}{C} ratios.
The SiO fundamental bands most likely have a photospheric origin in the 
all three stars.
\end{abstract}


\section{Introduction}

Novae are semi-detached binary systems in which a cool star fills
its Roche lobe. Material from the cool star (the secondary) flows
through the inner Lagrangian point and spirals onto the surface
of a white dwarf (WD) via an accretion disc (AD). In time, the base of 
the accreted layer becomes degenerate, and a thermonuclear 
runaway (TNR) occurs on the WD surface. This results in the explosive
ejection of up to $\sim10^{-4}$\Msun\ of material at several
100s to several 1000s of \vunit: a nova eruption has occurred.

After the eruption, mass transfer resumes and in time ($\ltsimeq10^6$~yrs),
conditions again become suitable for another nova eruption.
All novae are therefore recurrent, but those that repeat on a
human timescale ($\ltsimeq100$~yrs) are known as Recurrent 
Novae \citep[RNe; see][for reviews]{evans08,schaefer10}, as opposed to
the more leisurely Classical Novae (CNe).

While several hundred CNe are known, only about a
dozen RNe are known \citep[see][for a summary]{anupama08}.
A CN loses its ``classical'' status
if it is seen to erupt again within less than a few decades of eruption:
undoubtedly there are systems that are RNe, as normally defined,
masquerading as CNe
\citep[see, e.g.][and references therein]{pagnotta14}.
Indeed, it has been suggested
\citep{pagnotta14,della20} that as many as 30\% of CNe may be
recurrents.

RNe are of particular interest in that the WD in these systems
are thought to be close to the Chandrasekhar limit 
\citep*{starrfield16}, 
and potentially therefore progenitors of Type~Ia supernovae. 
The latter of course are a crucial tool in the determination 
of large-scale cosmic structure \citep{reiss98,perlmutter99}, 
so an understanding of RNe has wide application.

In this paper we report on the analysis of the optical, 
near-infrared (NIR) and mid-IR spectra of the RN \rnsgr, 
obtained while it was in quiescence, both before 
and after its 2019 eruption (see Section~\ref{3890}). 
A discussion of the NIR spectra 
obtained during the 2019 eruption will be presented elsewhere
\citep{evans22b}.

\section{The RN \rnsgr}
\label{3890}
The RN \rnsgr\ has undergone eruptions in 1962, 1990 
\citep[e.g.,][]{anupama08}, and most recently on 2019 Aug 27.87 
\citep{pereira}. The orbital period of \rnsgr\ is listed by 
\cite{schaefer09} as $519.7\pm0.3$~days, with semi-major axis 
362\Rsun. However, a more recent analysis, using extensive 
spectroscopic and photometric data \citep{mikolajewska21}, gives 
the orbital period as 747.6 days with binary separation
$a\sin{i}\simeq430$\Rsun, where $i$ is the binary inclination. 
\cite{mikolajewska21} found that $i\simeq68^\circ$, so the system
is not eclipsing. While these orbital parameters are 
consistent with a red giant (RG) that fills (or almost fills) its Roche lobe,
there seems to be no clear evidence in the visual light curve 
that can clearly be attributed to ellipsoidal variations 
\citep{mikolajewska21}. Indeed data from the OGLE survey 
\citep{soszynski13} suggest that the light variations of
\rnsgr\ can be quite erratic.

\cite{darnley12} have discussed the quiescent
optical and NIR colours of \rnsgr. Its optical properties
suggest a sub-giant secondary, while its NIR colours are more 
consistent with a RG secondary. \citeauthor{darnley12} suggest that 
this discrepancy indicates a high accretion rate and hence high 
luminosity AD, especially if the binary system is close to face-on;
however this conflicts with the inclination determined by 
\cite{mikolajewska21}. Its quiescent $K_s$ magnitude and orbital 
period place it together with RS~Oph \citep[although][use the 
\cite{schaefer09} period]{darnley12}.
Its rate of decline from maximum, and its high ejection velocities, 
during eruption \citep[see, e.g.,][]{evans22b} also place it in
\citeauthor{darnley12}'s ``RG-Novae'' category.

The mass ratio is $q=0.78\pm0.05$ for \rnsgr\ 
\citep[][]{mikolajewska21}. Using the dependence of $\log{g}$ on $q$
\citep[see Figure~1 of][]{pavlenko20b}, this translates to
$\log{g}=0.72\pm0.02$ for the RG in \rnsgr. We assume $\log{g}=0.7$ here, 
although the calculated spectral energy distribution (SED) 
is not sensitive to the value of $\log{g}$ assumed.

The secondary star in \rnsgr\ has been variously classified as M8~III
\citep{williams91} and M5~III \citep*{harrison93,anupama99};
the observations on which these classifications were
based were carried out before \citep{williams91} and
after \citep{harrison93,anupama99} the 1990 eruption.
The effective temperature corresponding to these classifications
ranges from $T_{\rm eff}\sim3100$~K to $\sim3400$~K
\citep[see, e.g.,][]{belle21}.

The reddening to \rnsgr\ has been discussed in detail by
\cite{page20}, who found that various authors have deduced 
$E(B-V)$ values in the range 0.46--1.1, depending on the method
used. \citeauthor{page20} also found that the ultraviolet (UV) 
data are best dereddened with a SMC extinction law
\citep{prevot84}, which is steeper in the UV than the 
``conventional'' Galactic extinction law and lacks the 
2175\AA\ extinction bump. The
SMC law is required to give a satisfactory account of the 
``2175\AA'' bump, and leads to a high UV flux.
A reddening value $E(B-V)=0.59$, based on an analysis of the 
interstellar \pion{Na}{i}~D lines from CHIRON spectra (see below),
was obtained by \cite{munari19}.
In view of the various uncertainties we shall treat the reddening
as a parameter to be determined as part of the fitting routine.
However we use a conventional Galactic extinction law
\citep{cardelli89}.

\begin{figure*}
\includegraphics[width=12cm]{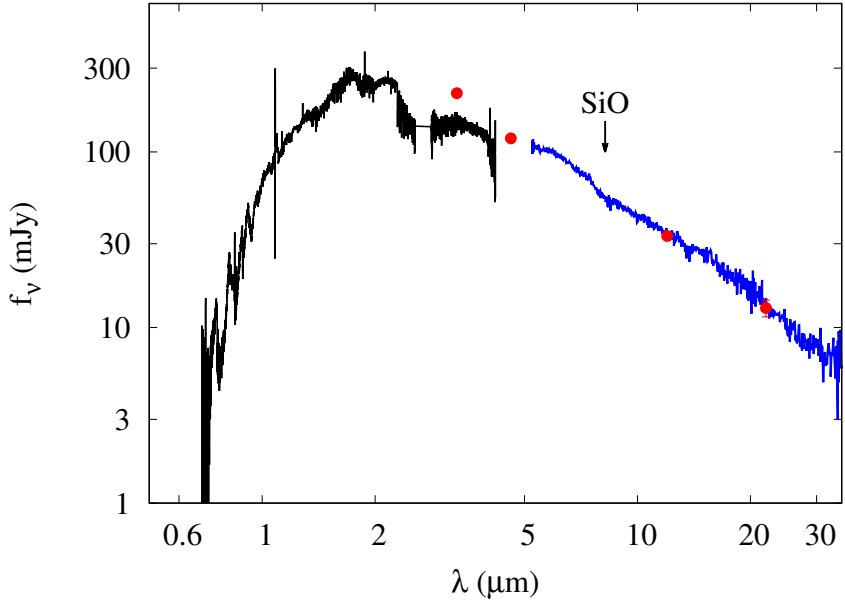}
\caption{Quiescent spectrum of \rnsgr, using post-2019-eruption 
ground-based data from 2020 June 6 (black), pre-eruption {\it Spitzer}
IRS data (blue; obtained in 2007 October) and pre-eruption
{\it WISE} survey data (red points; obtained in 2010 March).
The sharp feature just longward of 1\mic\ is \pion{He}{i} $\lambda=1.083$\mic,
that at 1.8\mic\ is Pa-$\alpha$.
The location of the SiO fundamental at 8\mic\ is indicated.
See text for details
\label{Spitzer}}
\end{figure*}

\section{Observations of V3890~Sgr in quiescence}
\rnsgr\ was observed both before and after the 2019 eruption, 
at optical, NIR and mid-IR wavelengths, covering the wavelength range
4070\AA--38\mic. None of the observations are contemporaneous. 
We summarise the observations below, and in Table~\ref{dates}.
In order to model the SED of the RG,
we must determine whether the RG photosphere is irradiated by 
the WD. Using the ephemeris of \cite{mikolajewska21}, we find 
the orbital phases given in Table~\ref{dates};
phase zero is when the RG component is at inferior conjunction.

\begin{table*}
\caption{\label{dates} Dates and orbital phases for the observations.}
\begin{tabular}{lcccccc}

Facility & Instrument  & Wavelength & Resolution & UT Date       & JD --    &  Orbital\\ 
         &             &  Range     &  $\lambda/\delta\lambda$ & YYYY-MM-DD.DD & 2450000   &  phase$^*$    \\\hline
\multicolumn{7}{c}{Before 2019 eruption} \\ \cline{2-6}  \noalign{\smallskip}
{\it Spitzer} & IRS    & 5--38\mic  &   $\sim100$   &  2007-10-11.27 & 4384.77  & 0.52 \\ \noalign{\smallskip}
{\it WISE}$^{**}$ & --- & 3.6--22\mic & $\sim1$ & 2010-03-27.60 & 5283.10 & 0.72 \\\noalign{\smallskip}
\multicolumn{1}{c}{$''$} & --- & 3.6--22\mic & $\sim1$ & 2010-09-25.85 & 5465.35 & 0.97 \\\noalign{\smallskip}
SMARTS   & CHIRON & 4071--8908\AA & 27800 & 2018-05-13.35 & 8251.85 & 0.69 \\ \noalign{\smallskip}
SMARTS        & CHIRON & 4071--8908\AA  & 27800 & 2018-05-16.32 & 8254.82 & 0.69 \\ \noalign{\smallskip}

\multicolumn{7}{c}{After 2019 eruption} \\\cline{2-6} \noalign{\smallskip}
SMARTS$^{\dag}$       & CHIRON & 4070 --  8907\AA & 78000 & 2019-08-29.06 & 8724.56 & 0.33 \\\noalign{\smallskip}
SMARTS$^{\dag}$       & CHIRON & 4070 --  8907\AA & 27800 & 2019-09-03.03 & 8730.53 & 0.34 \\\noalign{\smallskip}
SMARTS$^{\ddag}$       & CHIRON & 4070 --  8907\AA & 27800 & 2019-09-16.02 & 8742.52 & 0.35 \\\noalign{\smallskip}
SMARTS$^{\ddag}$       & CHIRON & 4070 --  8907\AA & 27800 & 2019-09-26.02 & 8752.52 & 0.36 \\\noalign{\smallskip}
IRTF    & SpeX   & 0.7--2.6\mic  & 750 & 2020-05-20.53 & 8990.00 & 0.68 \\ \noalign{\smallskip}
IRTF    & SpeX   & 0.7--4.2\mic  & 1200 & 2020-06-06.56 & 9007.06 & 0.70 \\ \noalign{\smallskip}
LBT        & PEPSI &  4219--7426\AA  &  130000    &  2020-06-10.27 & 9010.77 & 0.71 \\ \noalign{\smallskip}
IRTF    & SpeX  & 0.7--2.6\mic& 1200 & 2021-05-25.48  & 9360.02  & 0.18 \\ \noalign{\smallskip} 
IRTF    & SpeX   & 0.7--2.6\mic  & 1200 & 2021-07-07.40 & 9402.90 & 0.23 \\ \noalign{\smallskip}\hline\hline
\multicolumn{7}{l}{$^*$From the ephemeris of \cite{mikolajewska21}.
Phase zero is when the RG component is in front of the WD}\\
\multicolumn{7}{l}{$^{**}${\it WISE} photometry was taken over the period 2001 
March 27 -- September 26. Ssee text for details.}\\
\multicolumn{7}{l}{$\dag$These spectra were taken {\em close to the peak} of the 2019 eruption.}\\
\multicolumn{7}{l}{$\ddag$These spectra were taken during the decline of the 2019 eruption.}\\
\end{tabular}
\end{table*}

\subsection{Before: \sirtf\ IRS}
\label{irss}
\rnsgr\ was observed on 2007 October 11.27 UT with the \spitzer\ 
\citep{spitzer,spitzer-g} InfraRed Spectrograph \citep[IRS;][]{houck} 
as part of the \sirtf\ programme PID~40060 (P.I. A. Evans).
The spectrum was retrieved from the 
{\it Combined Atlas of Sources with \sirtf\ IRS Spectra}
\citep[CASSIS;][]{cassis}, and is shown in Fig.~\ref{Spitzer}.
The spectrum is almost featureless. The only possible feature
is the SiO fundamental in absorption (see Fig.~\ref{Spitzer}), 
which was also seen in the {\it Spitzer} IRS spectrum of the RNe T~CrB 
\citep{evans19} and RS~Oph \citep{rushton22}.
The \rnsgr\ Spitzer data were obtained at orbital phase 0.52, 
when the WD was in front of the RG; irradiation effects may therefore
be important in this case.

\subsection{WISE}

\rnsgr\ was detected in the 
Wide-field Infrared Survey Explorer survey \citep[{\it WISE};][]{WISE}
in 2010 March and September. {\it WISE} data for CNe and RNe were presented by 
\cite{evans14}. The fluxes for \rnsgr\ are listed in \citeauthor{evans14} 
but are given here (Table~\ref{wise}) for completeness, and plotted in 
Fig.~\ref{Spitzer}.

While \cite{evans14} stated that {\it WISE} data are consistent with 
a blackbody photosphere at 2310~K, with no evidence of line or
dust emission, the blackbody fit was not entirely satisfactory.
This is clearly because the {\it WISE} data cover only the 
Rayleigh-Jeans tail of the stellar photosphere and do not capture
the intricacies of the stellar emission (see Fig.~\ref{Spitzer}).

The {\it WISE} data were obtained over the period 2010
March 27 -- September 26. However the survey passes were concentrated in 
$\sim1$~day-long ``bursts'', around MJD 55282.5 and MJD 55464.0;
the mean MJD values are 55282.60 (2010 March 27.60: ``epoch~1'') 
and 55464.85 (2010 September 25.85: ``epoch~2''), the values entered in 
Table~\ref{dates}. The weighted mean {\it WISE} magnitudes for these two epochs
are $W1=7.842\pm0.025$, $W2=7.841\pm0.020$ (epoch~1)
and $W1=7.935\pm0.030$, $W2=7.952\pm0.023$ (epoch~2) respectively.
The mean values of $W1$ and $W2$ for these two epochs 
differ significantly at better than the 0.5\% level.
This difference can be attributed to the fact that epoch~1 data were
obtained at quadrature (when any ellipsoidal effect would have been at
its greatest extent), while epoch~2 data were obtained when the RG 
was in front of the WD (see Table~\ref{dates}, but also Section~\ref{3890}).

\begin{table}
\centering
\caption{\label{wise} Fluxes for \rnsgr\ from the {\it WISE} survey
\citep{evans14}.}
\begin{tabular}{cccc}
  {\it WISE} & $\lambda$ & Flux density  & Flux error \\ 
band &   (\mic)   &  (mJy)   & (mJy) \\ \hline
$W1$ &  3.3 & 216 & 5 \\
 $W2$ & 4.6 & 119 & 2 \\
 $W3$ & 12 & 33.2 & 0.6 \\
 $W4$ & 22 & 12.93 & 1.42 \\\hline
\end{tabular}
\end{table}

\subsection{Before and after: SMARTS}

\rnsgr\ was observed as part of the Stony Brook/SMARTS 
Nova program \citep{walter12}, on  2018 May 13.35 and 16.32 
(both UT), with the CHIRON echelle spectrograph. 
The spectra covered the wavelength range 4071\AA--8908\AA, 
at resolution $R=\lambda/\delta\lambda=27800$.  
In this paper we discuss mainly
the 2018 May 16 pre-2019 spectrum; 
similar data were also obtained on 2019 September 16.02 and 
26.02 (both UT), after the 2019 eruption. High 
resolution ($R=78000$) data, covering the same wavelength 
range, were obtained on 2019 August 29 close to the peak
of the 2019 eruption, and are also  discussed here.

The data were retrieved from the SMARTS Nova program
website\footnote{http://www.astro.sunysb.edu/fwalter/SMARTS/NovaAtlas}.
These data were obtained at a variety of orbital phases, as detailed
in Table~\ref{dates}.

\subsection{After: IRTF}

NIR spectra of \rnsgr\ were obtained on 2020 May 20.53
and 2021 May  25.48 UT with the infrared spectrograph SpeX 
\citep{rayner03} on the NASA Infrared Telescope Facilty IRTF. 
Full details of the data reduction are given in \cite{evans22b}
and are not repeated here. 
In each case the spectra showed essentially the ``normal'' 
SED of the RG component, with few emission lines superimposed. 
By these dates the system had 
essentially returned to quiescence \citep{woodward19}.
The spectrum for 2020 June 6 is included in Fig.~\ref{Spitzer}, to
give an overall impression of the SED in quiescence.

\begin{figure*}
   \centering
      \includegraphics[width=12cm]{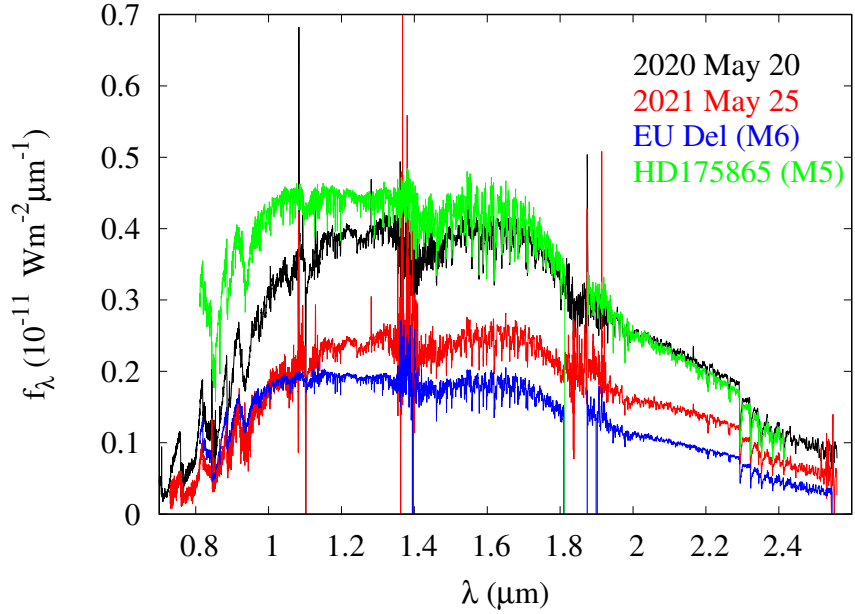}
  \caption{The two IRTF spectra obtained post outburst.
  Also shown are the spectra of EU~Del (M6~IIII) and HD175865 (M5~III)
  from the IRTF Spectral Library \citep{rayner09}.
  EU~Del and HD175865 have been reddened by $E(B-V)=0.4$ (see text).  
  \label{irrad-diff}}
\end{figure*}

As noted in Table~\ref{dates}, the post-eruption IRTF data were 
obtained at two distinct orbital phases. 
These two spectra are shown in Fig.~\ref{irrad-diff},
together with the spectra of two M giants 
(EU~Del, M6, and HD175865, M5) from the IRTF Spectral Library 
\citep*{rayner09}. The data for the two M giants have been
reddened by $E(B-V)=0.4$, to enable comparison with the observed
spectrum of \rnsgr.

\subsection{After: PEPSI}

High spectral resolution ($R = 120 000$) optical observations of \rnsgr\
in quiescence were obtained with the Potsdam Echelle Polarimetric and
Spectroscopic Instrument
\citep[PEPSI;][]{2015AN....336..324S} on the Large Binocular Telescope
\citep[LBT;][]{2012SPIE.8444E..1AH} on 2020 June 10.2653~UT with the
200\mic\ fibre using both apertures of the twin 8.4-m mirrors. 
Two sets of blue/red-arm cross-dispersing (CD) combinations were used, 
III/V (4800-5441~\AA/6278-7419~\AA) and II/IV (4265-4800~\AA/5441-6278~\AA), 
with an integration time for each pair of 40 minutes $\times2$ exposures. 
The average seeing, as measured from the peripheral wavefront sensors,
was 1\farcs01 and \rnsgr\ was near 2.34~airmasses at the start of the observation. 
The spectrophotometric standard 58~Aql was also observed with a similar 
instrumental setup. The raw CD data were processed through the
standard PEPSI pipeline (SDS4PEPSI version: 20210715), in which the barycentric
wavelengths were corrected for a $-8912.305$~m~s$^{-1}$ radial 
velocity to the barycenter of the Solar System. The various CD 
exposures were combined with a 
weighted average, and then flux normalized (dimensionless units).

\section{Modelling the RG spectrum}

\begin{figure*}
\includegraphics[width=12cm]{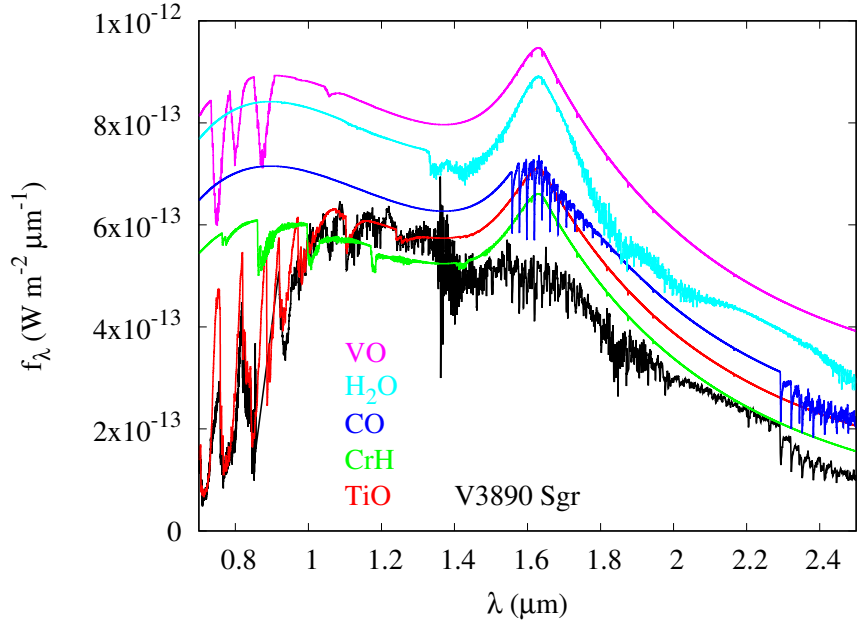}
\caption{Identification of the H$_2$O bands and the main diatomic molecules 
that contribute to the SED of the RG in \rnsgr; 
the individual molecular contributions
have been displaced vertically by arbitrary amounts for clarity.
Diatomics that have minimal effect on the spectrum (such as FeH) 
are not shown. The spectrum of \rnsgr\ is from 2020 May 20 and
has been dereddened.
\label{atoms1}}
\end{figure*}

To fit the observed SED over a wide spectral range we used 
classical model atmospheres computed in local thermodynamic
equilibrium \citep[LTE;][]{pavlenko13}. 
The contributions of selected diatomic molecules and H$_2$O
to the model SED in the wavelength range 0.7\mic--2.6\mic\ 
are shown in Fig.~\ref{atoms1},
in which each model curve includes the contributions
of the molecule and H$^-$ only. The H$^-$ is responsible for the peak
at $\sim1.6$\mic.

\subsection{Synthetic spectra and model atmospheres}

To generate synthetic spectra, we used the SAM12 model atmosphere code 
\citep[][and references therein]{pavlenko20a,pavlenko20b}. The procedure for 
computing synthetic spectra is described in \cite{evans19}, which also gives
sources for the molecular line lists. 
Line profiles were computed using Voigt profiles, and damping 
constants were taken from the line list databases, or computed in 
the \cite{unsold27} approximation. Synthetic spectra were computed in 
wavelength steps of 0.025\,\AA\ (0.025\AA\ to 1\AA\ for the Spitzer data). 
We determined the value of the turbulent velocity ($V_t$)
simultaneously with carbon isotope ratio for all individual spectra.
Also, our fitting procedure did not reveal any notable rotational 
velocities\footnote{Here $i$ is the inclination of the RG rotation axis,
not the orbital inclination.} $v\sin{i}$; 
therefore the theoretical spectra were convolved with a pure Gaussian
profile in order to model the instrumental broadening. 

We computed a grid of RG model atmospheres for the effective
temperatue range $T_{\rm eff}$ 2800--3700~K in steps of 100~K and
having solar \citep{asplund} abundances; $\log{g}$ was taken to
be 0.7 for all models. 

The best fit to the observed spectra was achieved  by the $\chi^2$
procedure described in \cite{pavlenko14}. We give a few 
details here to aid understanding of our procedure. 
As part of the fit, the function
\begin{equation}
S= \sum_{i=1}^{N} s_i^2
\end{equation}
is minimised, where 
$s_i= |F^{\rm obs}_i - F^{\rm comp}_i|$; $F^{\rm obs}_i$ 
and $F^{\rm comp}_i$ are the observed and computed fluxes, 
respectively, and $N$ is the number of the wavelengths points  
used in the minimisation procedure. Three parameters were used in 
our minimisation procedure: 
the heliocentric velocity of \rnsgr\ in \vunit, 
the flux scale normalisation parameter, and the full width at half 
maximum (FWHM) used for the smoothing Gaussian. 
They were determined for every fitted spectral range.
In our analysis we omitted some spectral ranges that contained 
artifacts provided by strong noise, telluric absorption, 
bad pixels, etc. The minimisation sum $S$ was  computed on a 
3D grid of radial velocity sets, flux normalisation factors, and 
FWHM parameters. Errors in the fit were evaluated as 
$\Delta S = \sum s_i/N$.

The data were dereddened by $E(B-V)=0.4, 0.5, 0.6$ 
\citep[cf.][]{mikolajewska21} prior to fitting. 

\section{Results}

\begin{figure*}
   \centering
      \includegraphics[width=12cm]{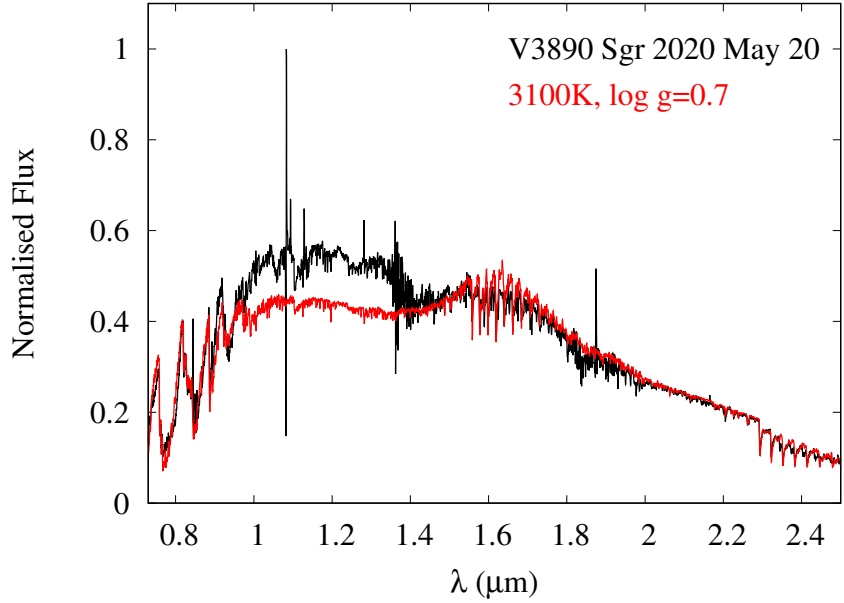}
  \caption{Model fits (red) to the dereddened 
  SED of V3890~Sgr (black) on 2020 May 20 (orbital phase 0.68). 
  The model atmosphere has $T_{\rm eff}=3100$~K, 
  $\log{g}=0.7$; solar abundances are assumed, except for $\mbox{[C]}=0.2$.
  The best fit is for $E(B-V) = 0.4$. 
  The flux scale has been normalised so that the maximum $f_\lambda$ in
  the observed spectrum has value unity. 
  \label{_fits}}
\end{figure*}

\begin{figure*}
\includegraphics[width=12cm]{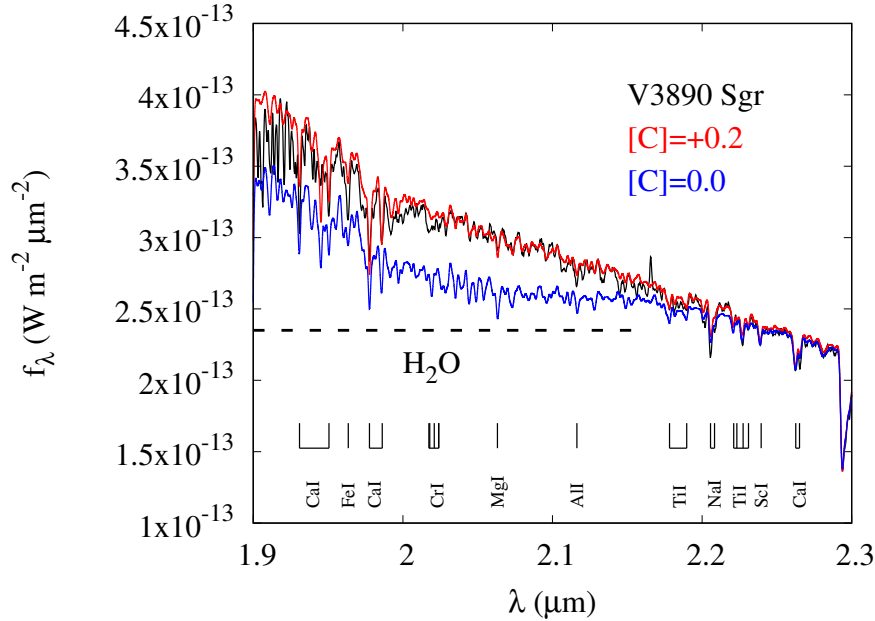}
\caption{Fit to the dereddened NIR SED of \rnsgr\ for 2020 May 20; 
atomic features are identified. 
Red curve is a synthetic spectrum with $T_{\rm eff}=3100$~K, $\log{g}=0.7$
and [C] = +0.2; blue curve has [C]=0.0. 
The broken line indicates the extent of the water band.
The first overtone \nucl{12}{C}O 
bandhead is just visible at 2.29\mic. \label{atoms2}}
\end{figure*}

\begin{figure*}
   \centering
      \includegraphics[width=12cm]{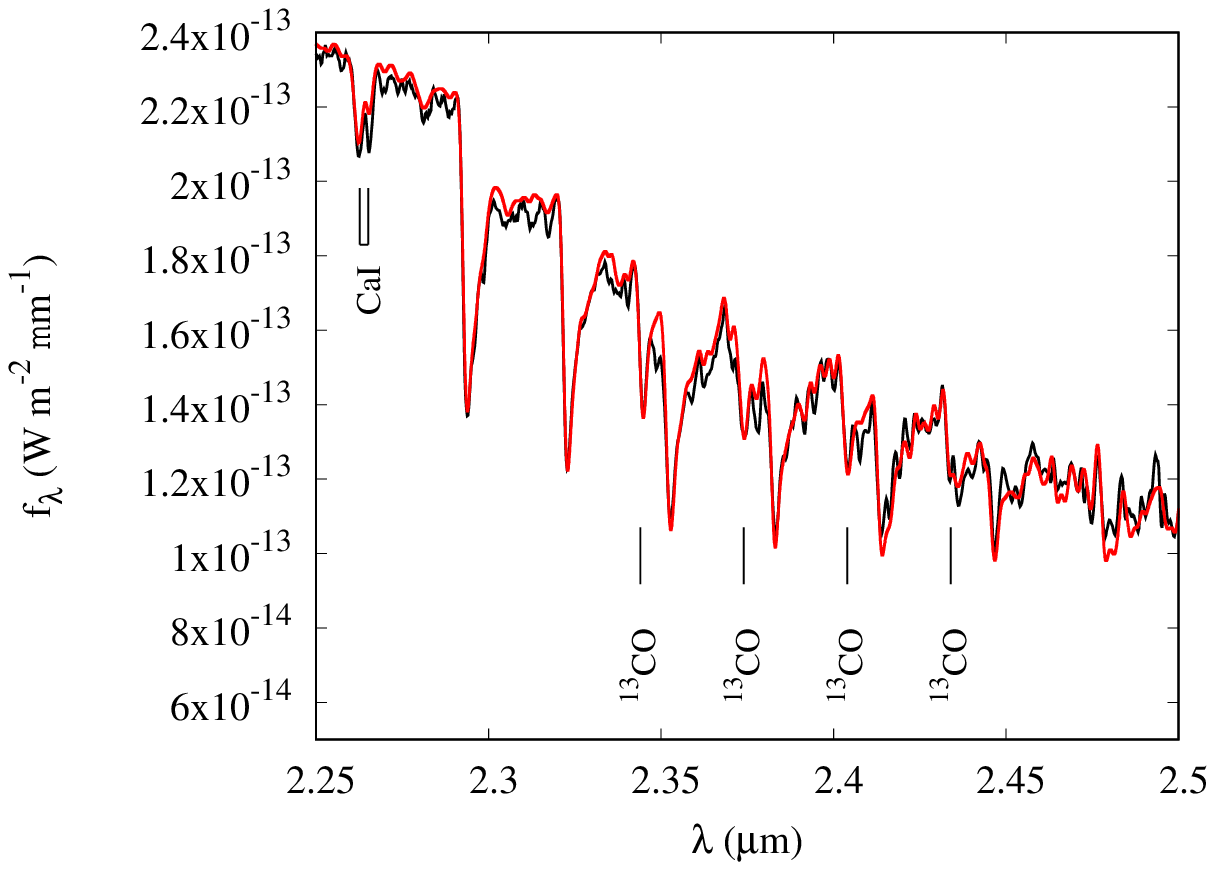}
  \caption{Best fit to the observed first overtone CO bands in \rnsgr.
  Data from 2020 May have been dereddened. The \nucl{13}{C}O band heads
  and \pion{Ca}{i} are identified.\label{co}}
\end{figure*}

\subsection{The IRTF data}

\subsubsection{SED, $T_{\rm eff}$ and abundances}
\label{Na-abund}
While our approach is simplified, we obtained good fits to 
the NIR spectra using solar abundances, with the 
exception of carbon (see Figs~\ref{_fits} and \ref{atoms2}). 
We adopt\footnote{$\mbox{[C]} =  
\mbox{[C/Fe]} = \mbox{(C/Fe)}_{\rm V3890~Sgr} - \mbox{(C/Fe)}_{\odot}$.}
$\mbox{[C]}=+0.20\pm0.05$~dex to reduce the effect of absorption by the water 
and TiO molecular bands, which are too strong if the solar abundance
of carbon is assumed (see Fig.~\ref{atoms2}). This is because the additional
carbon captures oxygen, resulting in a decrease in the abundance of H$_2$O 
and TiO. The only exception to the quality of the fit 
is in the 1.00--1.45\mic\ region, where we see a
significant over\-estimation of the opacity in the theoretical model
(note that the theoretical curve falls below the observed spectrum). 
All four IRTF spectra yield 
the same discrepancy in this wavelength region, even though the spectra
were all obtained approximately at quadrature (see Table~\ref{dates}). 
In general, it is found that theoretical fluxes are larger than 
observed due to a lack of opacity in the models. In the case of the 
IRTF data on \rnsgr, however, we see the opposite effect: in our model 
the opacity is {\em overestimated.} Alternatively the temperatures at 
the depths at which the 1.00--1.45\mic\ flux originates differ from our models.
This undoubtedly points to the limitations of the model used, and that 
a more sophisticated model than that used here is 
needed for such a complex object.

The values of $T_{\rm eff}$ and $E(B-V)$ for all four dates,
obtained from the fitting procedure, are given in 
Table~\ref{4sets}, and the best fit for 2020 May 20 is shown in
Fig.~\ref{_fits}. From Table~\ref{4sets} one may see that we obtained
two different values of $T_{\rm eff}$, namely 3000~K and 3100~K. 
These values do not differ within the uncertainty of 
$\pm200$~K, and both values were found for dates with similar orbital
phase ($\simeq0.7$ and $\simeq0.2$) so any difference cannot be due to 
irradiation effects. We adopt $T_{\rm eff}=3050\pm200$~K from fitting
the NIR data.

We obtained the same solution for the reddening ($E(B-V)=0.4$) for all
four dates, and this value is at the lower end of the range given 
in \cite{page20}. Our low value of $E(B-V)=0.4$ is based on fitting 
the SED, but according to \cite{mikolajewska21}, the SED of \rnsgr\
may be affected by additional flux from the AD. 
Such addional flux, which increases to the blue, provides partial 
compensation for the reddening, and leads to a lower $E(B-V)$ value.

There are a number of strong absorption lines in the NIR spectrum 
of \rnsgr. Some of the stronger atomic features are identified in 
Fig.~\ref{atoms2}. In general, the determination of atomic abundances 
by fitting the stronger lines in low resolution spectra is affected 
by a number of uncertainties (such as non-LTE, saturation of lines, etc.). 
Nevertheless we have determined the abundances of a number of species 
by fitting observed line profiles.

In general the abundances are close to solar, but there is an 
overabundance of sodium relative to solar, by a factor $\sim1.0\pm0.3$. 
Since the NIR \pion{Na}{i} lines used in the analysis are not
resonance lines, we are confident that they can not be interstellar.
Such an overabundance is not unexpected as a result
of the TNR \citep{starrfield09,starrfield20}. However, as discussed by
\cite{pavlenko20b}, the extent to which the TNR products pollute the RG
photosphere is unclear.
 
\subsubsection{The \nucl{12}{C}/\nucl{13}{C} ratio\label{coo}}

We have determined the \nucl{12}{C}/\nucl{13}{C} ratio
for four separate datasets after \rnsgr\ had returned to 
quiescence. These values, together with the determined values 
of $T_{\rm eff}$ and the microturbulent velocity $V_t$, 
are summarised in Table~\ref{4sets}.

The fit of our synthetic spectra across the first overtone CO bands
to the 2020 May IRTF spectrum  is shown in Fig.~\ref{co}; details of our 
procedure are given in \cite{pavlenko20b}.

\begin{table*}
 \centering
 \caption{Determination of the \nucl{12}{C}/\nucl{13}{C} from four separate datasets.
 $R$ is the effective spectral resolution.\label{4sets}}
 \begin{tabular}{ccccccc}
Date & $T_{\rm eff}$  & $V_t$  & \nucl{12}{C}/\nucl{13}{C} & $E(B-V)$ & $R$ & Orbital \\
YYYY-MM-DD     &   (K) &  (\vunit) & & &  & phase \\\hline
2020-05-20 &  3100 &   2.2   &   27   & 0.4 &    1050 & 0.68 \\
2020-06-06 &  3000 &   2.2   &   24   & 0.4 &    1300 & 0.70 \\
2021-05-25 &  3000 &   2.2   &   21   & 0.4 &   1300 & 0.18 \\
2021-07-07 &  3100 &   2.2   &   27   & 0.4 &    1300 & 0.23 \\   \hline  
 \end{tabular}
\end{table*}

\begin{table*}
 \centering
 \caption{The \nucl{12}{C}/\nucl{13}{C} ratio, dust temperatures, and SiO in 
 the RG components of RNe. \label{12-13}}
 \begin{tabular}{ccccl}
 RN & Dust & SiO & \nucl{12}{C}/\nucl{13}{C} & \multicolumn{1}{c}{Reference} \\ \hline
 RS Oph & $600\pm100$~K silicate &  &  & \cite{evans07b} \\
         & 400--500~K silicate            &   &      & \cite{rushton10} \\      
        &          &  & $16\pm3$          & \cite{pavlenko10} \\
        &        & Absorption: photospheric  &      & \cite{rushton22} \\  
 T CrB & None & Absorption: photospheric & $10\pm2$ & \cite{evans19}, but see  \\
   &  &  &  & \cite{evans22a} for important erratum \\
 \rnsgr & 400~K; composition  & Absorption: photospheric & $25\pm2$ & This work \\ 
  & indeterminate &  & & \\ \hline\hline
 \end{tabular}
\end{table*}

We have now determined the \nucl{12}{C}/\nucl{13}{C} 
ratio for the RG components of three RNe, as summarised in 
Table~\ref{12-13}. These values all seem to be consistent with 
that expected ($\sim20$) after first dredge-up, for stars with 
initial masses $\gtsimeq2$\Msun\ \citep[see, e.g.][]{karakas},
and are significantly lower than the solar value of $\sim90$. 
However as with the sodium abundance, the RG photosphere may be 
polluted by the products of the 1990 TNR and the measured 
\nucl{12}{C}/\nucl{13}{C} ratio in these systems may
reflect this process.

\subsubsection{SiO first overtone absorption\label{SiO1}}

\begin{figure*}
 \centering
 \includegraphics[width=12cm]{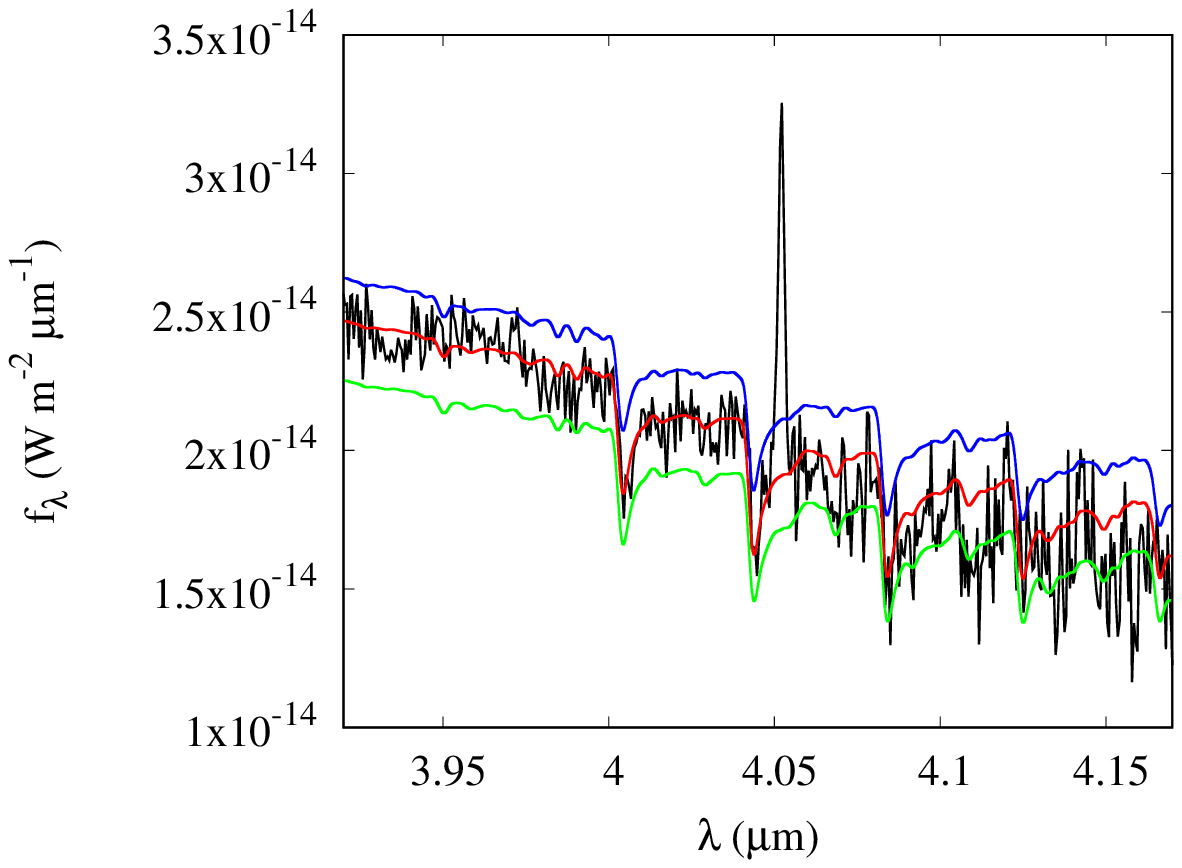}
 \caption{Fit of model atmospheres having various Si abundances
 to the first overtone SiO bands.
 Black curve: \rnsgr, dereddened as discussed in text.
 Red curve: computed first overtone SiO, assuming solar Si abundance.
 Blue curve: as red curve but with Si abundance $\mbox{[Si]}=-0.5$.
 Green curve: as red curve but with Si abundance $\mbox{[Si]}=+0.5$.
 Blue and green curves have been displaced upwards and downwards
 respectively for clarity.
 The emission feature is \pion{H}{i} Brackett $\alpha$ (5--4)
 at 4.052\mic.
 \label{SiO_4mic}}
\end{figure*}

The spectrum obtained on 2020 June 26 extends to 4.2\mic,
and clearly shows the presence of first overtone SiO in
absorption (see Fig.~\ref{SiO_4mic}). A fit of the model 
atmosphere already used to model the NIR data to this
spectral range suggests that the Si abundance is essentially solar
($\mbox{[Si]}=0.0\pm0.5$ in Fig.~\ref{SiO_4mic}); the uncertainty
in the abundance of Si is greater than it is for C because of the 
noisier data in the 4\mic\ band. This has implications
for the modelling of the SiO fundamental (see Section~\ref{sio_f}
below).

\subsection{The CHIRON and PEPSI optical spectra}

In this subsection we discuss the CHIRON and PEPSI spectra, and 
consider what they imply for the wind and environment of the RG.

\subsubsection{CHIRON}

\begin{figure*}
\centering
\includegraphics[width=7.cm]{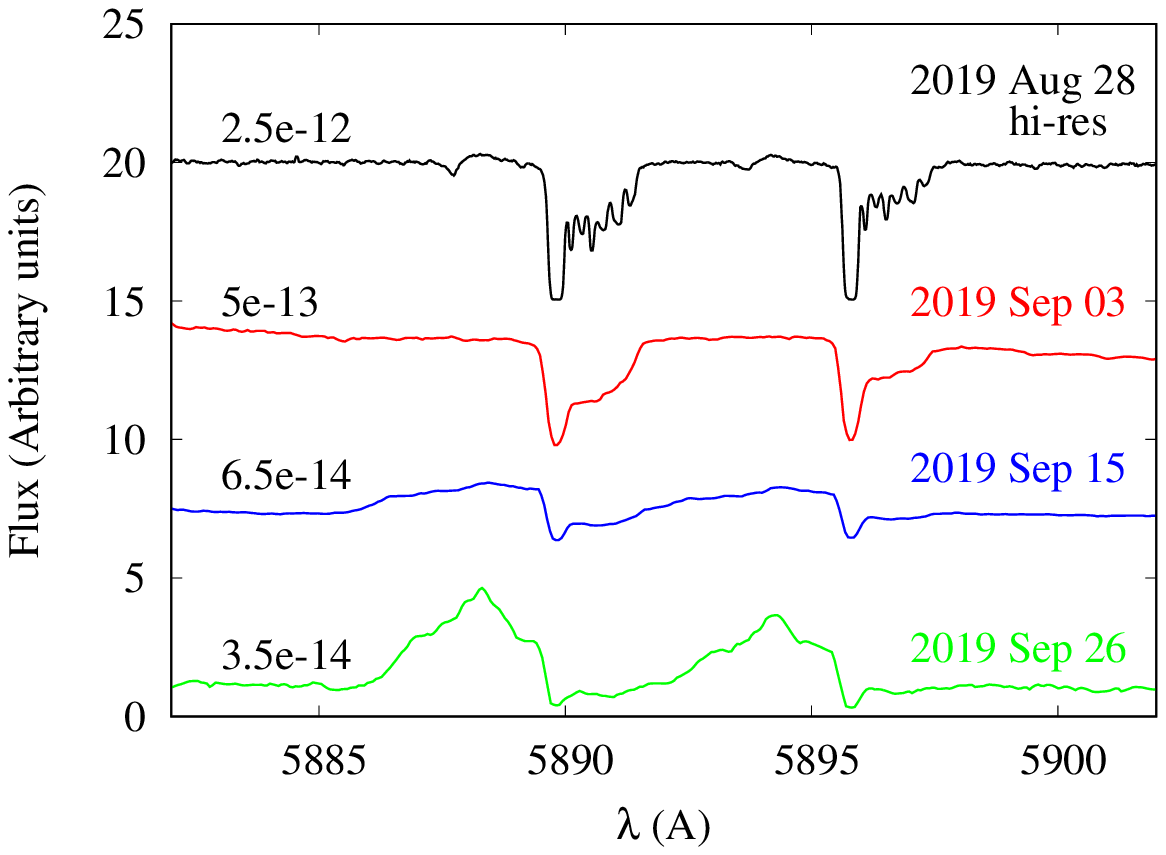}
\includegraphics[width=7.cm]{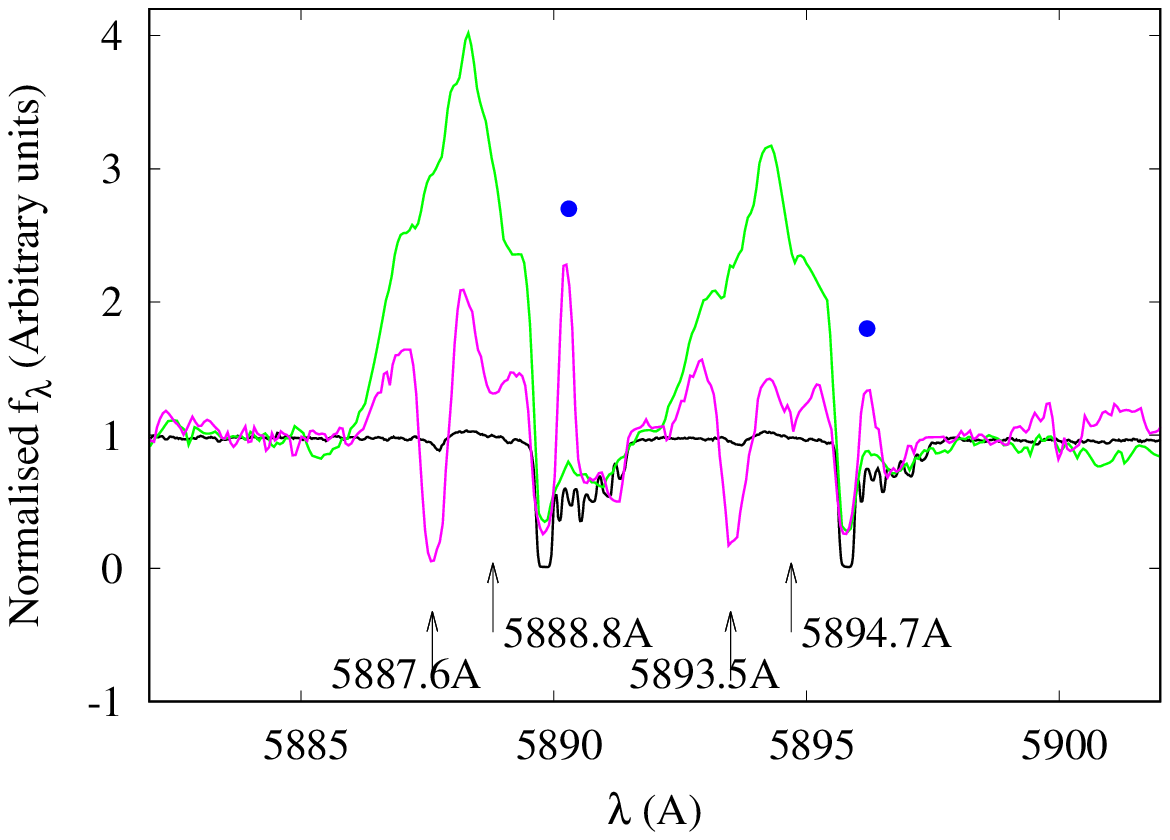}
\includegraphics[width=7.cm]{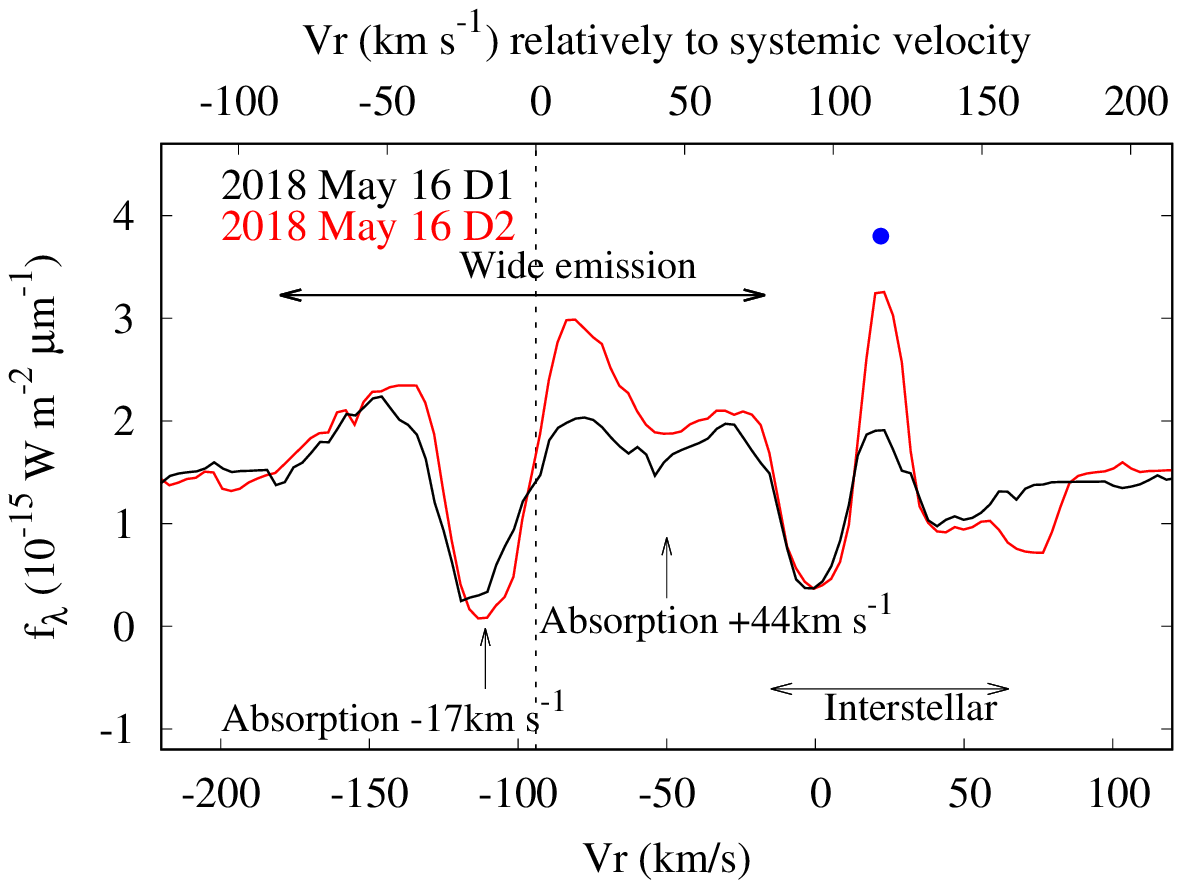}
 \includegraphics[width=7.5cm]{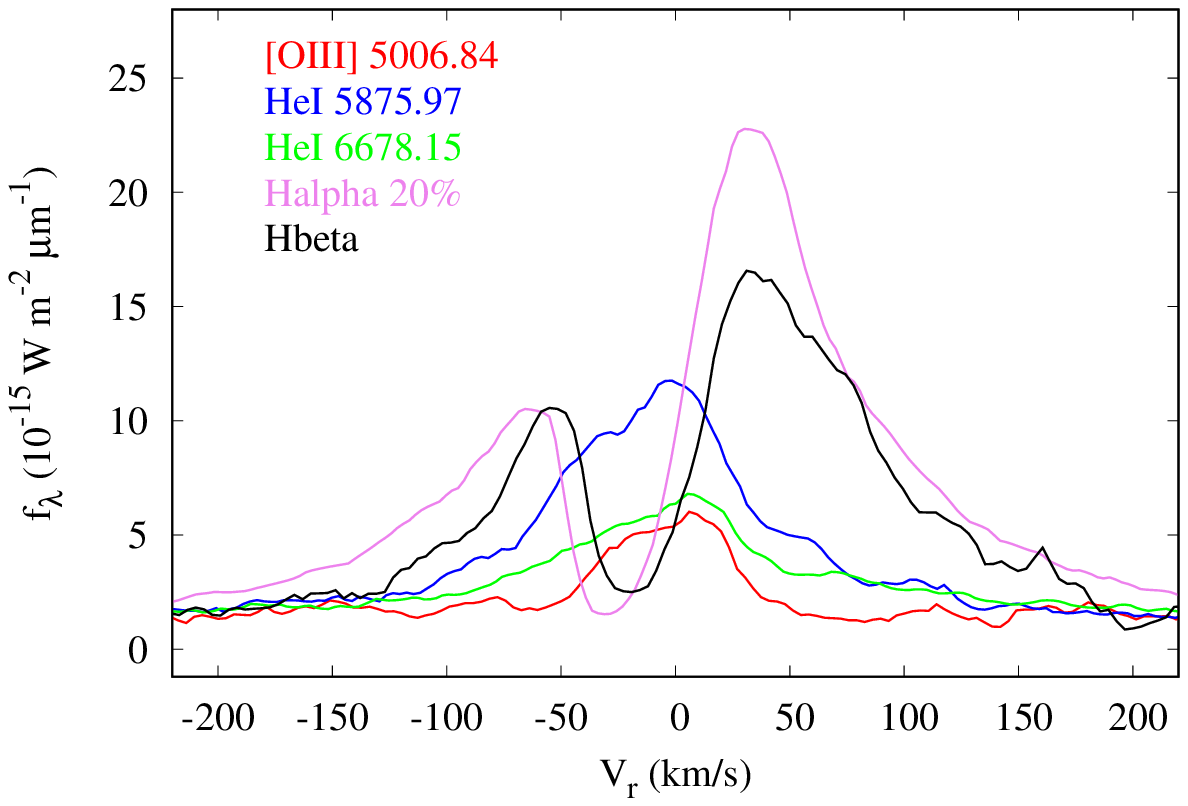}
\caption{Top left: high resolution spectra of \rnsgr\ in the region of 
the \pion{Na}{i}~D lines on the dates indicated. The numbers attached 
to each spectrum give the flux densities at 5883\AA.
Top right: post-2019 eruption spectra on the dates indicated;
the 2019 August 28 (black) and 2019 September 26 (green) spectra 
from the left panel are replicated here, with the same colour coding.
The 2018 May spectrum (magenta) was acquired before the 2019 eruption.
The spectra have been normalised to the same continuum, for ease of comparison. 
The blue dot denotes telluric emission.
Bottom left: 2018 May 16 spectrum in the region of the \pion{Na}{i} D lines,
plotted in velocity space. 
The systemic velocity \citep[--94\vunit;][]{mikolajewska21} is shown 
by the dashed line. The blue dot denotes telluric emission. See text for discussion.
Bottom right: permitted lines of \pion{He}{i} 5875\AA, 6678\AA, and the 
forbidden \pion{O}{iii} 5006\AA{} line before the 2019 eruption; 
compare with the middle panel of Fig.~\ref{pepsi}.
\label{NaD}}
\end{figure*}

One day after the 2019 eruption (2019 Aug 28), the flux has clearly 
increased significantly, but only interstellar \pion{Na}{i} D~lines
appear to be present in this region (Fig.~\ref{NaD} top left). 
At high resolution, at least six interstellar features are evident, 
all at positive radial velocity. 
The radial velocity of the main interstellar 
\pion{Na}{i} feature is the same as that of the interstellar
\pion{K}{i} resonance doublet.

The first post-outburst low resolution spectrum 
was obtained on 2019 September 3. While the flux has decreased by a 
factor of five, only the interstellar lines are present, 
shown in red in the top left panel of Fig.~\ref{NaD}. 
By 2019 September 15, the flux 
had decreased by a further factor of seven, and broad blue-shifted 
\pion{Na}{i} emission from the expanding ejecta appears.
Possibly this component was present, but obscured, at earlier 
times because of the high outburst flux. On 2019 September 26, the flux had 
decreased by a further factor of three, and the broad \pion{Na}{i} emission 
features had become relatively stronger.

In the top right panel of Fig.~\ref{NaD}, the black/green curves 
correspond to the black/green curves in the top left panel, but now the 
pre-outburst spectrum (magenta) from 2018 May is included. The flux 
level in the latter is some 23 times lower than the 2019 September
26 value (green). The wings of the broad \pion{Na}{i} emission 
features arising from the ejecta (green) are clearly present in the 
2018 May spectrum.
However, as shown in the bottom left panel of Fig.~\ref{NaD},
two blue-shifted absorption features are seen superimposed
on the broad emission. Of these, one component (the stronger) at
5887.6/5893.5\AA\ (D$_2$ and D$_1$ respectively) at radial velocity 
$-17$\vunit, the other at 5888.8/5894.7\AA\ has radial velocity 
$+44$\vunit. 
There are weak absorption features in the {\em post-outburst} high
resolution data that correspond to the 
stronger $-17$\vunit\ component, which we suggest is the wind of the RG.
P~Cygni profiles may be present as well.

In the bottom left panel of Fig.~\ref{NaD}, 
we show the \pion{Na}{i} D~lines for 2018 May 16 in velocity space.
These data were obtained 18~years after the 1990 outburst, and 
1~year before the 2019 outburst.
All the above-mentioned features are labelled in the bottom left panel
of Fig.~\ref{NaD}. Two absorption 
features have radial velocities $+44$\vunit\ and $-17$\vunit. 
These lines are weak one month after the (2019) outburst, 
but much stronger in the pre- and post-outburst PEPSI data.
The profiles of the \pion{Na}{i} D lines, and the changes
therein, are likely due to the complex interactions between the 
RG wind and material ejected in the 2019 and previous eruptions,
and material remaining from the common envelope phase of evolution
\citep*[see][]{booth16}.

The bottom right panel of Fig.~\ref{NaD} shows the
permitted lines of \pion{He}{i} 5875\AA, 6678\AA, and the 
forbidden \fion{O}{iii} 5006\AA{} line before the 2019 eruption.
This is included here for comparison with Fig.~\ref{pepsi} below.

\subsubsection{PEPSI}

\begin{figure*}
\centering
\begin{center}
\includegraphics[width=18cm]{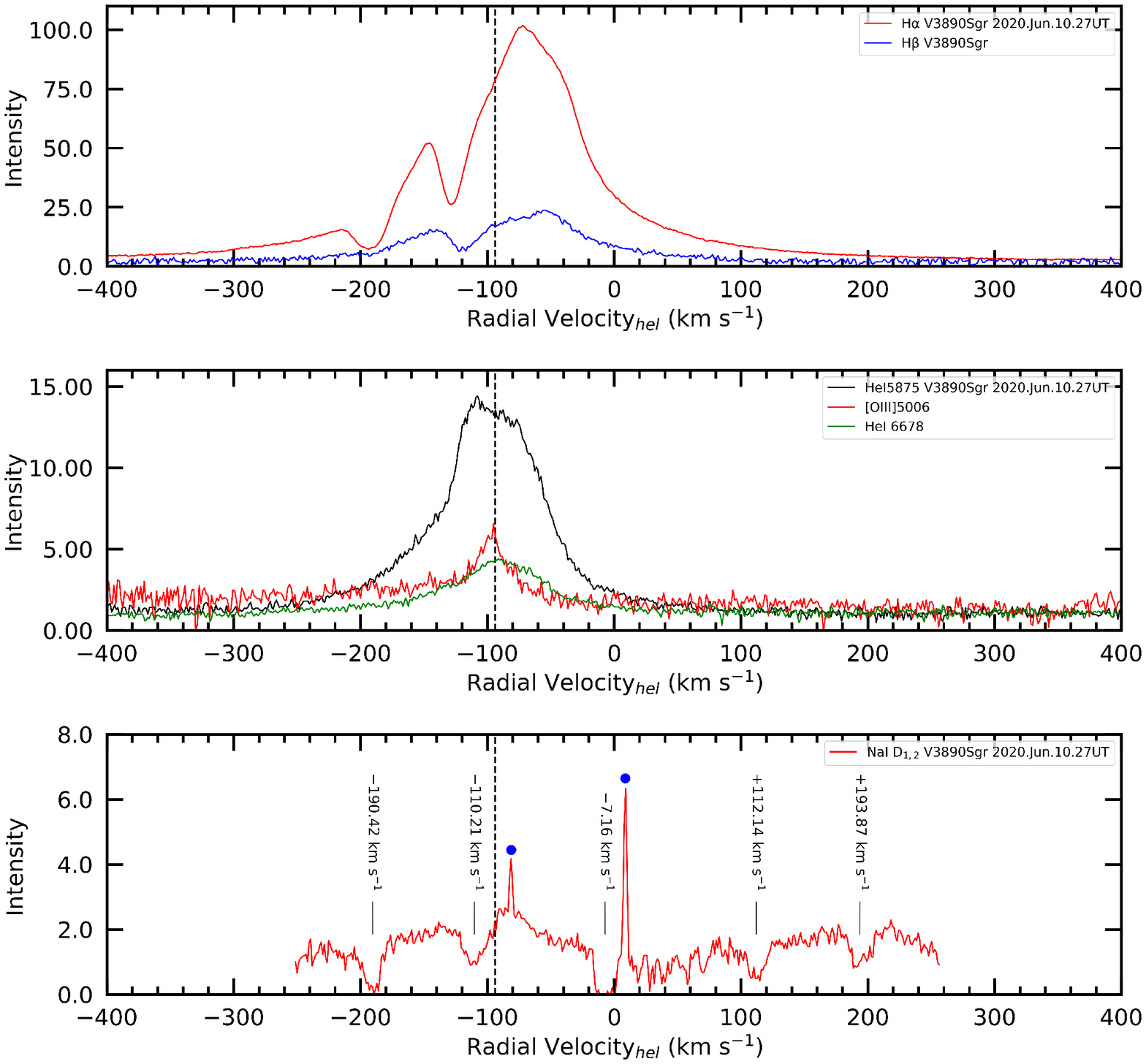}
\caption{ \label{pepsi} The PEPSI ($R=130,000$) observed velocity 
profiles of various strong emission lines in \rnsgr\ post the 2019 RN 
eruption at a phase of 0.709. Top: H$\alpha$ and H$\beta$. 
Middle: Permitted lines of \pion{He}{i} 5875\AA, 6678\AA, and the 
forbidden \pion{O}{iii} 5006\AA{} line.  
Bottom: The region of the \pion{Na}{i} D lines, for which the velocities 
are calculated with respect to the D$_{2}$ line (5889.995\AA{} air). 
The two narrow features marked by the blue circles are night sky sodium 
lines. Velocity components of various absorption features are indicated.
Note that there are only three distinct \pion{Na}{i} D absorption systems: 
the features at $-110.21$\vunit\ and $-190.47$\vunit\ are from the 
same systems as those at $+112$\vunit\ and $+193$\vunit, but for D$_1$.
The spectra are not absolutely flux calibrated, but are presented in 
continuum normalised intensity. These data have not been dereddened. 
The dashed vertical line in all the panels is the systemic velocity 
of the binary system, taken as $-94$\vunit{} \citep{mikolajewska21}.}
\end{center}
\end{figure*}

Segments of the 2020 Jun 10.27 UT spectra are presented in 
Fig.~\ref{pepsi}.
These data span a gap in the \cite{mikolajewska21} spectral coverage, 
being at a phase of  0.71, and were obtained with higher resolving power 
($R = 130,000$), enabling study of subtle structures in various line 
regions. The PEPSI spectra show a range of features, including emission 
lines of \pion{H}{i}, \pion{He}{i}, \fion{O}{iii}, \pion{O}{i}, and weaker 
emission from \pion{N}{ii} and \pion{Ca}{i}. The \pion{He}{ii} 4686\AA{} 
and the \fion{N}{ii} 6583\AA{} lines are not detected within the SNR-limit of
the spectrum. The \pion{Na}{i} D$_{1,2}$ region is complex, with many 
absorption components (but note the comment in the caption about the 
number of distinct absorption systems). The H$\alpha$ line is the 
strongest observed line with very broad wings, fit well by a 
Lorentzian with a double-peaked profile, the red peak being stronger 
than the blue peak (see Fig.~\ref{pepsi} top panel). On the other hand,
the H$\beta$ line resembles more a Gaussian in overall line profile shape, 
with a FWHM of $\sim165.18$\vunit. Both lines have absorption components 
of their blue wings, with the strongest component blue shifted 
$\sim-30$\vunit\ with respect to the binary systemic velocity 
\citep[94\vunit,][]{mikolajewska21}, while H$\alpha$ has another component
at $\sim100$\vunit. The PEPSI spectrum, 
together with the CHIRON spectrum from 2020 March 8,
clearly demonstrate that this 
feature was quite strong for at least 94 days, from 2020 March 8 
through 2022 Jun 10, before diminishing in intensity 13 days later 
\citep[see Fig.~7 in][]{mikolajewska21}. 

\cite{williams08}
reported transient absorption features, with velocities of
a few hundred \vunit, in high-resolution ($R=48000$)
optical spectra of CNe shortly after outburst. These features arise in
a wide range of heavy elements (e.g., Sc, Ti, V, Cr, Fe),
and progressively weaken with a few weeks of the eruption.
They suggest that
these features arise in material ejected by the secondary, and which
pre-dates the nova eruption. The behaviour of the H$\beta$
absorptions in \rnsgr\ is reminiscent of that described by
\citeauthor{williams08}, and may arise in the same way.

The broad \pion{He}{i} 5875\AA{}, 
\pion{He}{i} 6678\AA{} profiles are symmetric about the systemic 
velocity, while the forbidden emission line profile \fion{O}{iii} 
5006\AA{} line is sharply peaked with a FWHM $\sim35.28$\vunit{}
(see Fig.~\ref{pepsi} middle panel).

The permitted \pion{O}{i} 5014\AA{} and adjacent \pion{He}{i}
5016\AA{} features are symmetric, with similar integrated line 
intensities with FWHM $\sim83.89$\vunit\ and a narrower
FWHM $\sim56.62$\vunit, respectively. The \pion{Fe}{ii} 5018\AA{} 
line, seen in other symbiotic spectra \citep{quiroga02}, is not 
detected. The ratio of the integrated \fion{O}{iii} to \pion{O}{i} 
intensity is $\sim1.58$.

\subsection{The \sirtf\ IRS spectrum\label{irs}}

\subsubsection{Emission lines}

While there were strong emission lines (including coronal, 
fine structure and recombination) in the IRS spectrum of RS~Oph,
in data taken {\em after} its 2006 eruption \citep{evans07a,evans07b},
there are no emission lines in the {\it Spitzer} IRS spectrum
of the RN T~CrB in quiescence \citep{evans19}. 
\begin{figure}
   \centering
      \includegraphics[width=8cm]{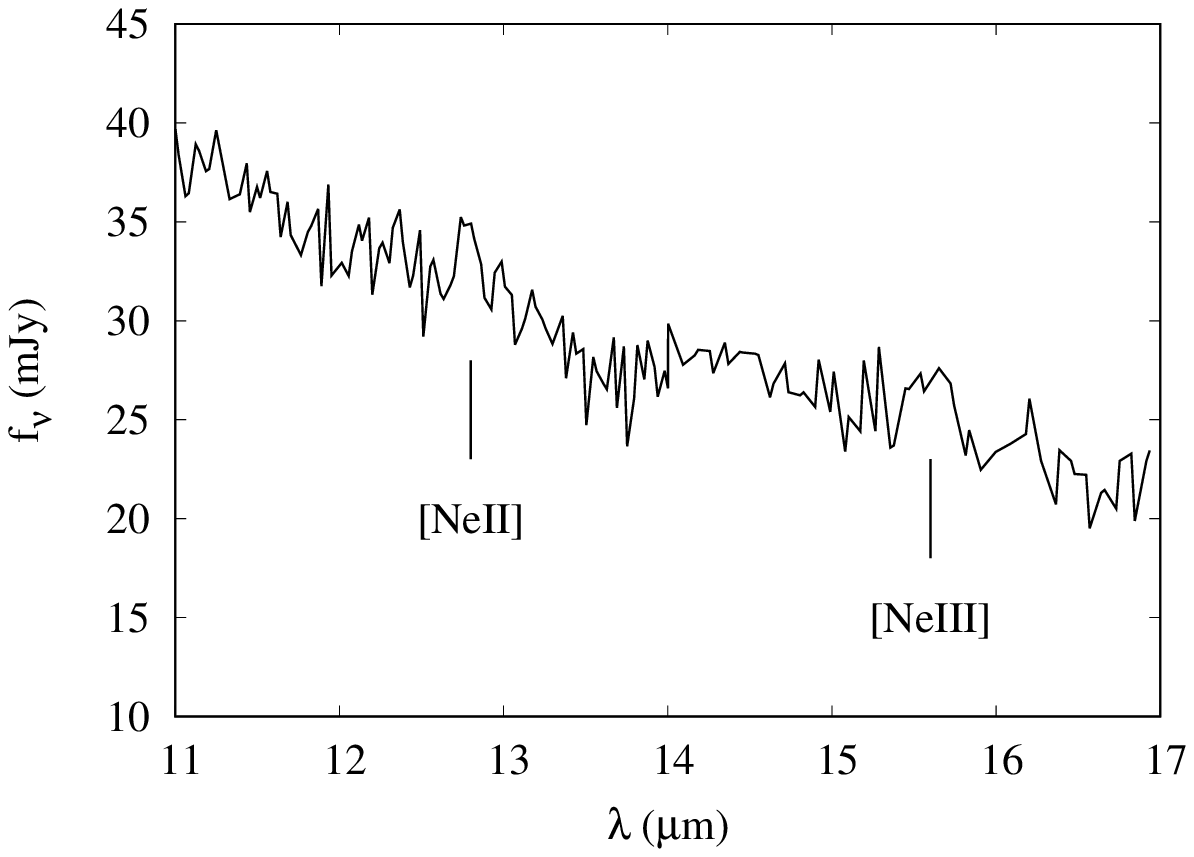}  
      \caption{Possible detection of neon fine structure lines
      in the {\it Spitzer} IRS spectrum of \rnsgr.
   \label{Nelines}}
\end{figure}

A segment of the {\it Spitzer} IRS spectrum of \rnsgr, obtained 
in 2007 October, is shown in Fig.~\ref{Nelines}. It shows marginal 
evidence for two emission lines, namely the \fion{Ne}{ii} 12.813\mic\
(flux $8.4[\pm2.3]\times10^{-18}$~W~m$^{-2}$) and 
\fion{Ne}{iii} 15.555\mic\ (flux $8.1[\pm2.3]\times10^{-18}$~W~m$^{-2}$)
fine structure lines. The flux ratio is 
\[ \frac{f(\mbox{\fion{Ne}{ii})}}{f(\mbox{\fion{Ne}{iii}})} \simeq 1.0\pm0.2\:\:.\]
We assume that the upper levels of the Ne transitions are excited
by electron collision. We take the effective collision strengths from 
the Iron Project\footnote{http://cdsweb.u-strasbg.fr/tipbase/home.html} \citep{badnell06}, and the ionisation equilibrium data for neon from 
\cite{arnaud85}. We determine the temperature of the gas in
which the neon lines originate by an iterative process, and find that
$T(\mbox{Ne})\simeq6.3\times10^4$~K. This does not seem unreasonable
for a gas that is ionised by a WD and an AD.

There is no clear evidence for the presence of the \fion{O}{iv} 
fine structure line at 25.89\mic\ 
\citep[ubiquitous in CNe;][]{helton12}, and present 
in the spectrum of RS~Oph \citep{evans07a,evans07b}, to a $3\sigma$ upper limit of 
$3.7\times10^{-18}$~W~m$^{-2}$. At $10^4$~K, the critical electron 
density above which the upper level of the \fion{O}{iv} 25.89\mic\ 
line is collisionally de-excited is $9.9\times10^3$~cm$^{-3}$, 
implying that the electron density in the gas from which the 
\fion{O}{iv} line would originate in the environment of 
\rnsgr\ exceeds this value.

\subsubsection{The SED}

In Section~\ref{Na-abund} we found $T_{\rm eff}=3050\pm200$~K 
from fitting the NIR data. There were no NIR observations of \rnsgr\ at the time
of the Spitzer observation so we have no value for $T_{\rm eff}$.
However, as this spectral region is on the Rayleigh-Jeans 
tail of the RG's photospheric emission, the fit is not sensitive to 
$T_{\rm eff}$. However the abundances ($\mbox{[C]}=+0.2$, 
$\mbox{[Na]}=+1.0$, $\mbox{[Si]}=+0.0$) we determined in the 
fits to the NIR data will be the same at the time of the Spitzer 
observations and we adopt these here. We again assume $\log{g}=0.7$,
but explore different values of $T_{\rm eff}$.

Around $T_{\rm eff}=3000-3100$~K (close to the value we deduced for 
the NIR), the water bands are too strong. They may be decreased by 
further increasing [C] to +0.25, but this is not an option 
as this strongly affects the SED in the NIR (see Fig.~\ref{atoms2} for
the effect of the water bands): the C abundance is constrained by
our fit to the NIR data. The alternative is to increase 
$T_{\rm eff}$, because at $T_{\rm eff}=3300$~K the water bands
disappear completely.

The fit to the IRS spectrum of \rnsgr\ is shown in the left panel of 
Fig.~\ref{_sio}, in which an excess flux relative to the RG photosphere 
(red curve) is evident longward of 7\mic. This additional flux is most 
plausibly provided by a dusty envelope around the RG, as reported by
\cite{rushton10,rushton14,rushton22} for the case of RS~Oph.
However in the case of RS~Oph, the dust shows strong silicate emission
\citep{evans07b,woodward08,rushton22}, which is not present in \rnsgr. 
We follow the scheme of \citeauthor{rushton22} and add a contribution 
from the dust, with temperature $T_{\rm dust}=400$~K, to the photospheric 
flux ($T_{\rm eff}=3300$~K). The additional blackbody emission greatly 
improves the fit, as shown by the blue curve in the left panel of 
Fig.~\ref{_sio}. 

We find that the model spectrum with $T_{\rm eff}=3400$~K resembles that for
$T_{\rm eff}=3300$~K, but for this case $T_{\rm dust}=450$~K: 
there is some ``degeneracy'' in the $T_{\rm eff}$ and $T_{\rm dust}$ values.
For example, we have the same fit for a 3400~K:450~K photosphere/dust 
temperature mix as we do for 3300~K:400~K, etc. However higher dust 
temperatures provide a greater contribution at shorter wavelengths. 
The fact that we see no evidence for dust in the NIR data sets limits on
$T_{\rm dust}$ and $T_{\rm eff}$. 

The value of $T_{\rm eff}$ required to fit the Spitzer 
data is somewhat higher than that obtained from fitting the NIR 
data (3050~K). The absence of the water bands, and the 
requirement that the carbon abundance must be +0.2~dex to fit 
the NIR data, limits $T_{\rm eff}$ at the low end to
3300~K. On the other hand, 
there is no evidence for dust in the NIR data, limiting $T_{\rm eff}$ to 
3400~K at high end. Thus, despite the fact that the SED is not sensitive 
to $T_{\rm eff}$ in the Spitzer spectral range, our value of $T_{\rm eff}$ 
is quite robust. Noting that the Spitzer data were obtained
when WD was in front of the RG, while all the NIR data were taken at
quadrature, it is likely that we see real changes in temperature 
(see Section~\ref{irss}). A similar effect was found in our analysis
of RS~Oph \citep{pavlenko16}, in which the overall effect of irradiation 
was to increase $T_{\rm eff}$ by 100--200~K.

\subsubsection{The SiO fundamental band}
\label{sio_f}
\begin{figure*}
   \centering
   \includegraphics[width=8cm]{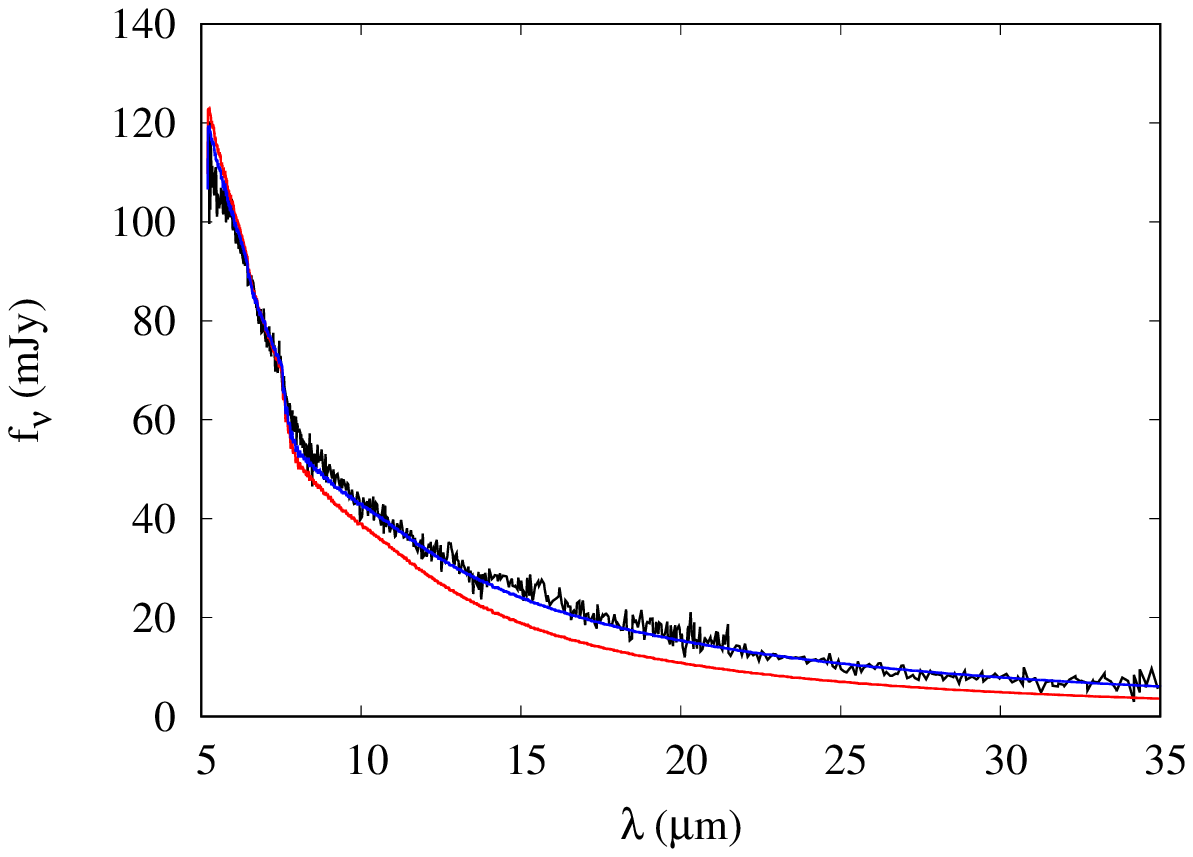}
         \includegraphics[width=8cm]{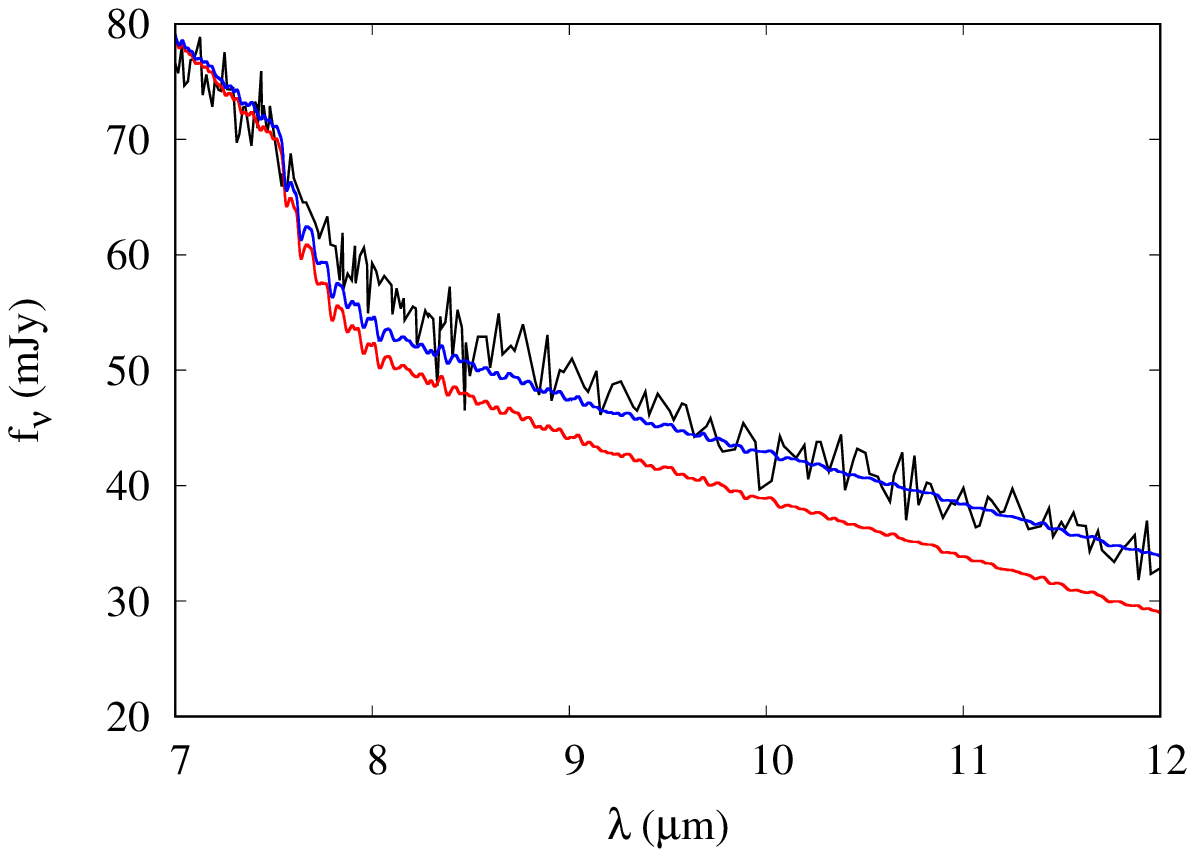}
  \caption{Left: Fit to the {\it Spitzer} IRS spectrum in the 
  5--35\mic\ region with a $T_{\rm eff}=3300$~K, $\log{g}=0.7$ 
  photosphere. Black: \rnsgr,
  Red: $T_{\rm eff}=3300$~K, $\log{g}=0.7$ photosphere, and
  solar Si isotopic ratios.
  Blue: as red but including dust at $T=400$~K.
  Right: Detail of fit in the 7--12\mic\ region; colour coding as left. 
  See text for details.
   \label{_sio}}
\end{figure*}

The presence of the SiO fundamental band near 8\mic\ in the spectrum of T~CrB 
was reported by \cite{evans19}. The Spitzer IRS spectrum of \rnsgr, 
shown in Fig.~\ref{_sio}, resembles that of T CrB in that there is no 9.7\mic\
silicate emission feature \citep[as there is in the case of the RN RS~Oph;][]
{evans07b, woodward08, rushton14, rushton22} but there is weak absorption
in the SiO fundamental. As in the case of T~CrB, this likely arises in the 
RG photosphere \citep{evans22a}.

The lines of the various SiO isotopologue fundamental transitions are strongly
blended, and the resolution of the Spitzer IRS data is too low to identify the 
individual features. In general, including the higher mass isotoplogues 
provides a stronger 8\mic\ SiO band. We have no information about the Si isotopic
ratios in \rnsgr: high resolution spectroscopy in the 4\mic\ first overtone
is required to robustly determine the Si abundance ratios. We have
therefore computed synthetic spectra on the assumption of solar Si isotopic
abundances. As in case of modeling the IRS SED, the SiO fundamental band shows 
a weak dependence on the adopted $T_{\rm eff}$ in our temperature range. 

The fit to the SiO fundamental is shown in Fig.~\ref{_sio}.
At $T_{\rm eff}=3300$~K, the 8\mic\ band is somewhat weaker 
than it is at 3100~K, but still stronger than it is in the observed spectrum.

The properties of the circumstellar dust, and of the absorptions in the 
SiO fundamental transition, are summarised in Table~\ref{12-13}; note that, 
in \cite{evans19}, the fit to the SiO fundamental was incorrect; see erratum
in \cite{evans22a}.

\section{Conclusions}

We have presented optical and infrared
spectroscopy of the recurrent nova \rnsgr, covering the 
wavelength range 4070\AA--38\mic. We find that
\begin{enumerate}
\item the surface gravity of the red giant component is
$\log{g}=0.72\pm0.02$, and its effective temperature is in the range
$T_{\rm eff}=3050\pm$200~K to $T_{\rm eff}=3300\pm$200~K; 
\item there is an overabundance of both
carbon (0.2~dex) and sodium (1~dex) relative to the solar value;
\item the \nucl{12}{C}/\nucl{13}{C} ratio $=25\pm2$, 
a value similar to that found in red giants in other recurrent 
novae;
\item the interpretation of the quiescent spectrum in the 
5--38\mic\ region requires the presence of photospheric SiO 
absorption and of cool ($\sim400$~K) dust in the red giant 
environment;
\item the region of the \pion{Na}{i} D lines includes
a number of interstellar components. There is also
evidence for interaction between ejecta from the 2019 eruption
and material accumulated in the plane of the binary.
\item there is evidence for two neon fine structure lines, namely 
\fion{Ne}{ii} 12.813\mic\ and \fion{Ne}{iii} 15.555\mic, the flux 
ratio of which implies an electron temperature of $6.3\times10^4$~K;
\item the three RG in RN systems we have considered thus far have
in common an overabundance of carbon and similar \nucl{12}{C}/\nucl{13}{C} 
ratios; both properties are typical of RG photospheres after first 
dredge-up in stars of low initial mass. All three also
have photospheric SiO in absorption.
\end{enumerate}

\section*{Acknowledgments}

We thank the referee for their helpful and supportive comments.

AE thanks John Southworth for a helpful discussion on the orbital phase.

This study was funded as part of the routine financing
programme for institutes of the National Academy of Sciences
of Ukraine. YP gratefully acknowledges the hospitality of the
Nicolaus Copernicus Astronomical Center, Toru\'n, and support from 
grant 2017/27/B/ST9/01128 financed by the Polish National Science Center,
while this work was being completed.
DPKB is supported by a CSIR Emeritus Scientist grant-in-aid and 
is being hosted by the Physical Research Laboratory, Ahmedabad.
RDG was supported by the National Aeronautics and Space Administration 
(NASA) and the United States Air Force (USAF).
FMW acknowledges support of the US taxpayers
through NSF grant 1611443.

The work is based in part on archival data obtained using 
the Infrared Telescope Facility operated by the University of Hawaii 
under a cooperative agreement with NASA.  

Data for this paper were obtained under IRTF programme 2020A-010.
The Infrared Telescope Facility is operated by the
University of Hawaii under contract 80HGTR19D0030 with the 
National Aeronautics and Space Administration.

Access to the SMARTS partnership is made
possible in part by research support from Stony Brook University.

The LBT is an international collaboration among institutions in the
United States, Italy and Germany. LBT Corporation partners are: The
University of Arizona on behalf of the Arizona university system;
Istituto Nazionale di Astrofisica, Italy; LBT
Beteiligungsgesellschaft, Germany, representing the Max-Planck
Society, the Astrophysical Institute Potsdam, and Heidelberg
University; The Ohio State University, The University of Notre Dame,
University of Minnesota and University of Virginia.

The authors would like to thank the 
SAO/NASA ADS team for the development and support
of this system.


\section*{Data availability}
The data used in this paper are available in the various observatory archives,
as follows:
\begin{enumerate}
 \item Spitzer: https://cassis.sirtf.com/atlas/
 \item WISE: http://wise.ssl.berkeley.edu/
 \item SMARTS: http://www.astro.sunysb.edu/fwalter /SMARTS/NovaAtlas/
 \item PEPSI: The PEPSI data underlying this paper will be shared on a reasonable request to the corresponding author.
 \item IRTF: http://irtfweb.ifa.hawaii.edu/research/irtf\_data \_archive.php;
\end{enumerate}

\bsp

\label{lastpage}

\end{document}